\documentclass[a4paper,twocolumn,aps,prl,superscriptaddress]{revtex4}
\usepackage{amsmath}
\usepackage{graphicx}
\usepackage{amssymb}
\usepackage{color}

\newcommand{\ket}[1]{\left|#1\right>}
\newcommand{\bra}[1]{\left<#1\right|}
\newcommand{\abs}[1]{\left|#1\right|}
\newcommand{\braket}[2]{\left<#1|#2\right>}

\begin{document}

\title{Fractional Quantum Hall State in Coupled Cavities}

\author{Jaeyoon Cho}

\affiliation{Department of Physics and Astronomy, University College London, Gower St., London WC1E 6BT, UK}

\affiliation{Centre for Quantum Technologies, National University of Singapore,
2 Science Drive 3, Singapore 117542}

\author{Dimitris G. Angelakis}

\affiliation{Centre for Quantum Technologies, National University of Singapore,
2 Science Drive 3, Singapore 117542}

\affiliation{Science Department, Technical University of Crete, Chania, Crete,
Greece, 73100}

\author{Sougato Bose}

\affiliation{Department of Physics and Astronomy, University College London, Gower St., London WC1E 6BT, UK}

\affiliation{Centre for Quantum Technologies, National University of Singapore,
2 Science Drive 3, Singapore 117542}

\date{\today}

\begin{abstract}
We propose a scheme to realize the fractional quantum Hall system with atoms confined in a two-dimensional array of coupled cavities. Our scheme is based on simple optical manipulation of atomic internal states and inter-cavity hopping of virtually excited photons. It is shown that as well as the fractional quantum Hall system, any system of hard-core bosons on a lattice in the presence of an arbitrary Abelian vector potential can be simulated solely by controlling the phases of constantly applied lasers. The scheme, for the first time, exploits the core advantage of coupled cavity simulations, namely the individual addressability of the components and also brings the gauge potential into such simulations as well as the simple optical creation of particles. 
\end{abstract}

\maketitle

The achievement of trapping ultracold atomic gases in a strongly correlated regime has prompted an interest in mimicking various condensed matter systems, thereby allowing one to tackle such complex systems in unprecedented ways~\cite{lewenstein07}.
A major class of simulable systems, distinct from the Hubbard model and spin systems, is that in a gauge potential, such as the fractional quantum Hall (FQH) system. The FQH effect arises when a two-dimensional (2D) electron gas is in the presence of a strong perpendicular magnetic field at a low temperature. The Hall resistance of such a system exhibits plateaus when the Landau filling factor $\nu$ takes simple
rational values~\cite{tsui82}. The FQH effect at fundamental filling factors $\nu=1/m$ for odd integers $m$ (even
integers for bosons) is accounted for by Laughlin's trial wave function (in the symmetric gauge)~\cite{laughlin83}\begin{equation}
\Psi_{m}(\{z_{j}\})=e^{-\frac{1}{4}\sum_{j}\abs{z_{j}}^{2}}\prod_{j<k}(z_{j}-z_{k})^{m},\label{eq:laughlin}\end{equation}
where $z_{j}=x_{j}+iy_{j}$ is the 2D position of the $j$th electron
in unit of the magnetic length $l_{B}\equiv\sqrt{\hbar/eB}$ with
$B$ being the magnetic field. The elementary excitation of this state
is a quasihole (quasiparticle), which has a fractional charge $+e/m$
($-e/m$) and obeys the anyonic statistics~\cite{arovas84}. 
To simulate such a system in trapped atoms, a major challenge is to create an artificial magnetic field as the atoms in consideration have no real charge. This is done with considerable difficulties, for instance, by rapidly rotating the harmonic
trap~\cite{wilkin00}, by exploiting electromagnetically induced
transparency~\cite{juzeliunas04}, or by modulating the optical
lattice potential~\cite{jaksch03,sorensen05}. Additionally, FQH systems are also simulable in Josephson junction arrays~\cite{stern94}.

Recently, coupled cavity arrays (CCAs)~\cite{hartmann06,rossini07,hartmann07} have emerged as a fascinating alternative for simulating quantum many-body phenomena, supported by diverse technologies, such as microwave stripline resonators, photonic crystal defects, microtoroidal cavity arrays, and so forth~\cite{hennessy07,majer07,aoki06}. CCAs have complementary advantages over optical lattices, such as arbitrary many-body geometries and individual addressability~\cite{cho07}. Recently, theoretical works have shown that the Mott-superfluid phase transition of polaritons~\cite{hartmann06,rossini07} and the Heisenberg spin chains~\cite{hartmann07} can be realized in CCAs. 
These works, however, relied only on globally addressing lasers and thus could not highlight the key advantage of CCAs, namely, the individual addressability in the sense that already they can be done similarly or better in optical lattices~\cite{jaksch98}. Moreover, simulating altogether distinct classes of systems such as those of itinerant particles in a gauge potential still remains open and this will be especially arresting if the particles themselves can be created by a purely optical means.
In this paper, we bring the Abelian gauge potential into the realm of many-body simulations using CCAs. We achieve this by actively exploiting the individual addressability, which eventually enables great versatility which has not been attainable in optical lattices. 

As a concrete example, we introduce a way of simulating FQH systems in CCAs. To be more specific, we consider a FQH system of bosonic particles confined in a 2D square lattice of spacing $a$ in the presence of a perpendicular and uniform artificial magnetic field $B$. Non-interacting free bosons in a single Bloch band are described by
the Hamiltonian
\begin{equation}
H_{0}=-t\sum_{\left<j,k\right>}c_{j}^{\dagger}c_{k}\exp\left(-i\frac{2\pi}{\Phi_{0}}\int_{j}^{k}\mathbf{A}(\mathbf{r})\cdot
d\mathbf{l}\right),
\label{eq:gauge}
\end{equation} where $c_{j}$
denotes a boson annihilation operator at site $j$ and
$\Phi_{0}\equiv h/e$ is the magnetic flux quantum. The summation
runs over nearest neighbor pairs. If we take a Landau gauge, this Hamiltonian is written
as\begin{equation}
H_{0}=-t\sum_{p,q}\left(c_{p+1,q}^{\dagger}c_{p,q}e^{-i2\pi\alpha
q}+c_{p,q+1}^{\dagger}c_{p,q}+h.c.\right),\label{eq:hamil}\end{equation}
where the positions of lattice sites are represented by
$a(p\hat{x}+q\hat{y})$ with $a$
being the spacing of the lattice. Here, $\alpha\equiv Ba^{2}/\Phi_{0}$,
the number of magnetic flux quanta through a lattice cell, plays a
crucial role in characterizing the energy spectrum, whose
self-similar structure is known as the Hofstadter
butterfly~\cite{hofstadter76}. In addition to this non-interacting
Hamiltonian, we consider a hard-core interaction between bosons,
which limits the number of particles that can occupy one site to a
maximum of one. In this limit, if we also take a continuum limit
$\alpha\ll1$, the Laughlin state~\eqref{eq:laughlin} is a very
accurate variational ground state, where the filling factor $\nu$
corresponds to the ratio of the number of bosons to the number of
magnetic flux quanta.

In order to realize the above situation, we consider a two-dimensional array of coupled cavities each confining a single atom with two ground levels, which will be representing an $s=\frac{1}{2}$ spin. First, we notify that, aside
from the additional phases, the Hamiltonian~\eqref{eq:hamil} in the
hard-core limit corresponds to that of an $s=\frac{1}{2}$ spin
lattice system with XX interaction, where the creation-annihilation
operation of the zero and one boson states is analogous to the spin
flip operation of the spin-down and up states. This
natural realization of the hard-core limit is contrary to the case
of optical lattices, wherein it is achieved in the limit of strong
on-site repulsion~\cite{jaksch03,sorensen05}. Moreover, as will be
seen later, every aspect of the system is optically controlled: bosons are created by simple optical pulses and the phases in the Hamiltonian are adjusted
simply by controlling the phases of applied lasers. This
optical control of the system would greatly simplify the
experiments, compared to the previous schemes involving mechanical
modulations of the system. Although in this work we mainly consider the
FQH systems, another great advantage is that unlike the previous schemes for optical lattices
any Abelian vector potential on a lattice can be also simulated simply by
adjusting the laser phases in accordance with the
formula~\eqref{eq:gauge}. The creation of a quasiexcitation, which is achieved by adiabatically inserting a flux quantum through an infinitely thin magnetic solenoid
piercing the 2D plane~\cite{laughlin83}, again reduces to the matter
of adiabatically changing the laser phases accordingly. It can be moved along the lattice cells by modulating the
laser phases, which would be useful for testing the fractional
statistics.

\begin{figure}
\includegraphics[height=0.83in]{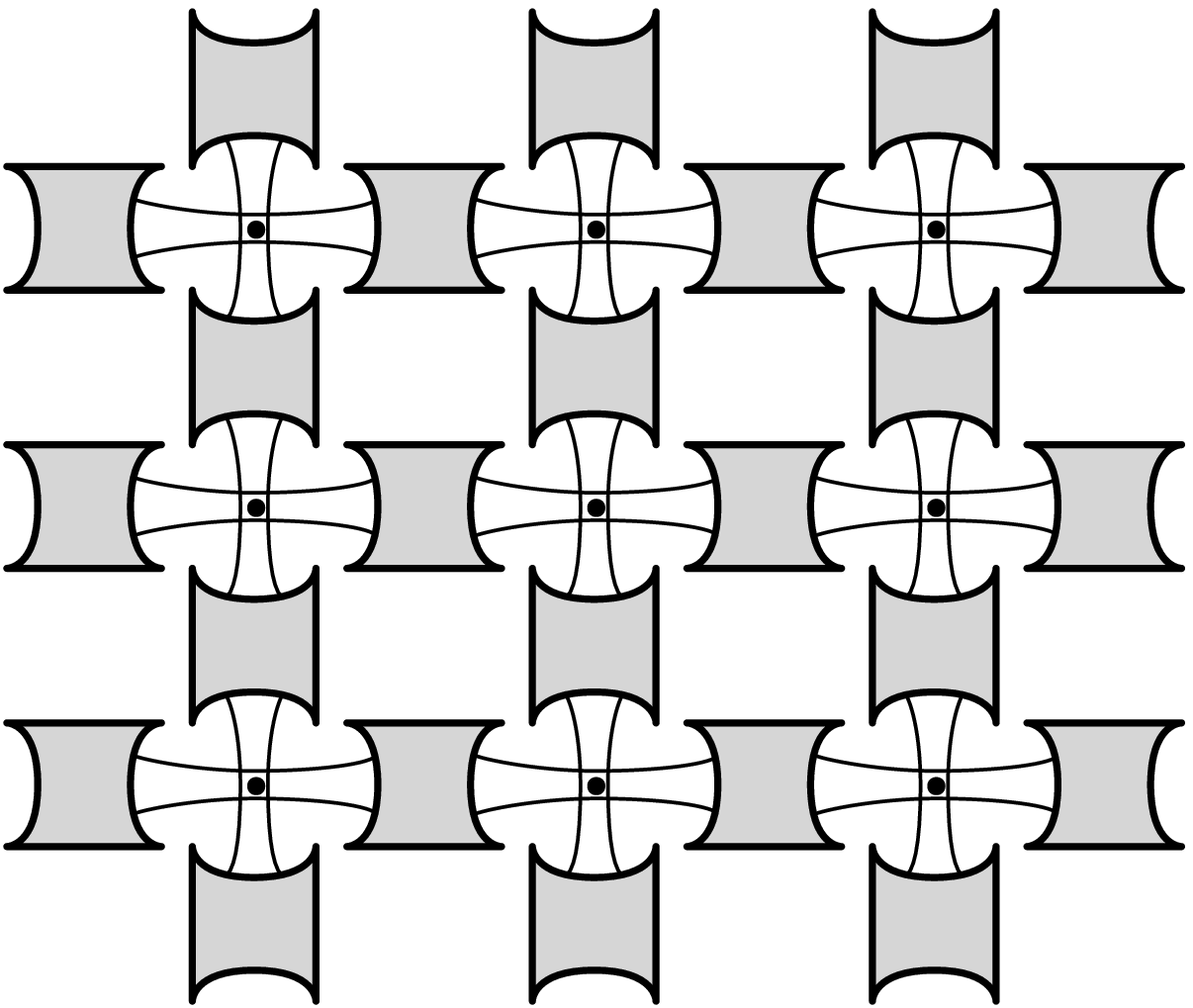}\caption{Schematic representation of a two-dimensional array of coupled cavities.
Each atom is confined in the intersection of two orthogonal cavity
modes, which are adjusted to have different resonant frequencies.}
\label{fig:cavity}
\end{figure}

Schemes for realizing the spin exchange Hamiltonian in an array of
coupled cavities have been established in recent papers~\cite{hartmann07}.
In these schemes, the spin exchange is mediated by inter-cavity hopping
of virtually excited cavity photons. An important difference in the
present case is that the spin exchanges are associated with phase changes depending on their locations and directions, which obviously cannot be excluded by local phase transformations.
For this reason, we introduce an asymmetry in the 2D geometry of 
coupled cavities, as shown schematically in FIG.~\ref{fig:cavity},
where two orthogonal cavity modes along the $\hat{x}$ and $\hat{y}$
directions have different resonant frequencies. Realizing this geometry
would be viable in several promising models for coupled cavities,
such as photonic bandgap microcavities~\cite{hennessy07} and superconducting
microwave cavities~\cite{majer07}. We assume the frequency difference
between the two modes is much larger than the atom-cavity coupling
rates. In this way, either direction of the spin exchange can be accessed
individually by choosing the laser frequency. We note, however, that
the above asymmetry is, in fact, not essential for our purpose. For
example, an array of microtoroidal cavities~\cite{aoki06}, in which
case realizing such a geometry is nontrivial,
can be also used by involving more lasers in the scheme. We discuss
this point later.

\begin{figure}
\includegraphics[height=0.83in]{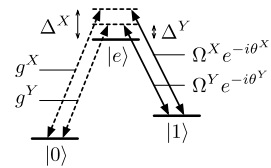}\caption{Involved atomic levels and transitions. There are two independent
Raman transitions mediated by an excited level $\ket{e}$ between
two ground levels $\ket{0}$ and $\ket{1}$, represented by superscripts
$X$ and $Y$, respectively.}
\label{fig:atom}
\end{figure}

We consider a simple atomic level and transition scheme
as shown in FIG.~\ref{fig:atom}. The atom has two ground levels
$\ket0$ and $\ket1$, and an excited level $\ket e$. There are two
cavity modes along the $\hat{x}$ and $\hat{y}$ directions, whose
annihilation operators are denoted by $a^{X}$ and $a^{Y}$, respectively.
The atom interacts with these cavity modes with coupling rates $g^{X}$
and $g^{Y}$, and with detunings $\Delta^{X}$ and $\Delta^{Y}$,
respectively. Two classical fields with (complex) Rabi frequencies
$\Omega^{X}e^{-i\theta^{X}}$ and $\Omega^{Y}e^{-i\theta^{Y}}$ are
applied, respectively, as in the figure. In the rotating frame, the
Hamiltonian reads
\begin{equation}
\begin{split}
H= & \sum_{\mu=X,Y}\sum_{j=(p,q)}\left[g^{\mu}e^{-i\Delta^{\mu}t}a_{j}^{\mu}(\ket{e}\bra{0})_{j}\right.\\
 & \quad\quad\quad\quad\quad\;\;\left.+\Omega^{\mu}e^{-i\theta_{j}^{\mu}}e^{-i\Delta^{\mu}t}(\ket{e}\bra{1})_{j}+h.c.\right]\\
 & -\sum_{p,q}\left(J^{X}a_{p+1,q}^{X\dagger}a_{p,q}^{X}+J^{Y}a_{p,q+1}^{Y\dagger}a_{p,q}^{Y}+h.c.\right),\end{split}
\end{equation}
where $J^{X}$ ($J^{Y}$) denotes the inter-cavity hopping rate of
the photon along the $\hat{x}$ ($\hat{y}$) direction, and the subscript
$(p,q)$ represents the cavity site. As mentioned above, we assume
$\Delta^{X}-\Delta^{Y}\gg g^{X},g^{Y}$, and also assume $\Delta^{\mu}\gg g^{\mu}\gg\Omega^{\mu},J^{\mu}$.
This requires the strong atom-cavity coupling in that $g^{\mu}\gg J^{\mu}$.
In this regime, the atomic excitation is suppressed, and adiabatic
elimination leads to an effective Hamiltonian
\begin{equation}
\begin{split}
H= & \sum_{\mu=X,Y}\sum_{j=(p,q)}\left[\delta^{\mu}a_{j}^{\mu\dagger}a_{j}^{\mu}(\ket{0}\bra{0})_{j}\right.\\
 & \quad\quad\quad\quad\quad\;\;\left.+\omega^{\mu}\left(e^{i\theta_{j}^{\mu}}a_{j}^{\mu}\sigma_{j}^{+}+h.c.\right)\right]\\
 & -\sum_{p,q}\left(J^{X}a_{p+1,q}^{X\dagger}a_{p,q}^{X}+J^{Y}a_{p,q+1}^{Y\dagger}a_{p,q}^{Y}+h.c.\right),\end{split}
\label{eq:noexc}\end{equation}
where $\delta^{\mu}=(g^{\mu})^{2}/\Delta^{\mu}$, $\omega^{\mu}=g^{\mu}\Omega^{\mu}/\Delta^{\mu}$,
and $\sigma^{+}=\ket{1}\bra{0}$. Here, we have ignored the ac Stark
shift induced by classical fields, which is negligible in our regime
(or it may be compensated by other lasers). Again, we assume $\delta^{\mu}\gg J^{\mu}\gg\omega^{\mu}$,
which can be satisfied, along with the above condition, when\begin{equation}
g^{\mu}/\Delta^{\mu}\gg J^{\mu}/g^{\mu}\gg\Omega^{\mu}/\Delta^{\mu}.\label{eq:cond}\end{equation}
In this regime, the cavity photon is suppressed, and adiabatic elimination
can be applied once more. We extend the method in Ref.~\cite{james07}
to keep up to the third order terms and take only the subspace with
no cavity photon. The effective Hamiltonian, in the rotating frame,
can then be derived as 
\begin{equation}
\begin{split}
H=-t\sum_{p,q} & \left[\sigma_{p+1,q}^{+}\sigma_{p,q}^{-}e^{i(\theta_{p+1,q}^{X}-\theta_{p,q}^{X})}\right.\\
 & \left.+\sigma_{p,q+1}^{+}\sigma_{p,q}^{-}e^{i(\theta_{p,q+1}^{Y}-\theta_{p,q}^{Y})}+h.c.\right],\end{split}
\label{eq:final}\end{equation}
where the parameters are chosen such that $t=J^{X}({\omega^{X}}/{\delta^{X}})^{2}=J^{Y}({\omega^{Y}}/{\delta^{Y}})^{2}$.
It is easy to see that this Hamiltonian reduces to the Hamiltonian~\eqref{eq:gauge}
if we adjust the phases of the classical fields as
\begin{equation}
\begin{split}
\theta_{p,q}^{X}= & \frac{2\pi}{\Phi_{0}}\int_{0,q}^{p,q}\mathbf{A}(\mathbf{r})\cdot d\mathbf{l},\\
\theta_{p,q}^{Y}= & \frac{2\pi}{\Phi_{0}}\int_{p,0}^{p,q}\mathbf{A}(\mathbf{r})\cdot d\mathbf{l}.\end{split}
\label{eq:generalphase}\end{equation}
The FQH Hamiltonian~\eqref{eq:hamil} is obtained if we adjust these
phases as $\theta_{p,q}^{X}=-pq\cdot2\pi\alpha$ and $\theta_{p,q}^{Y}=0$.
Note that the classical fields for $\theta_{p,q}^{Y}$ can
be replaced by one global field.

In order to check the validity of the adiabatic approximation from
Hamiltonian~\eqref{eq:noexc} to \eqref{eq:final},
we have performed a direct numerical diagonalization of Hamiltonian~\eqref{eq:noexc}.
We take a set of parameters $\delta^{\mu}/10=10\omega^{\mu}=J^{\mu}$,
which corresponds to a case where $\Delta^{\mu}/1000=g^{\mu}/100=\Omega^{\mu}=J^{\mu}$.
To eliminate the edge effects within a limited computational capability,
we consider a periodic boundary condition (i.e., a torus).
We consider a $4\times4$ lattice with $\alpha=1/4$ and two bosons,
hence four flux quanta in total and the filling factor $\nu=1/2$.
In view of the fact that the cavity photon is suppressed, we restrict
our calculation to the subspace wherein the maximum number of excitations
in a cavity is limited to one, i.e., $\langle a_{p,q}^{X\dagger}a_{p,q}^{X}+a_{p,q}^{Y\dagger}a_{p,q}^{Y}+\left(\ket{1}\bra{1}\right)_{p,q}\rangle\leq1$.
Up to the modification due to the torus geometry and a different gauge~\cite{haldane85a},
the ground state should be close to the Laughlin state~\eqref{eq:laughlin}
with $m=2$. From our numerical diagonalization, the fidelity between
the Laughlin state $\ket{\Psi_{2}}$ and the numerical ground state
$\ket{\Psi_{G}}$ is found to be $F_{G}=\abs{\braket{\Psi_{2}}{\Psi_{G}}}^{2}=0.976$.
We note that when the ideal Hamiltonian~\eqref{eq:hamil} is diagonalized
for the same $4\times4$ lattice, the fidelity of the ground state
is found to be $0.989$. The fidelity $F_{G}$ converges to this value
as $\delta^{\mu}/J^{\mu}$ and $J^{\mu}/\omega^{\mu}$ increase, which,
however, demands more strong atom-cavity coupling. Note also that
the non-unit fidelity is partly due to the finite $\alpha$, which
makes the effect of the lattice geometry non-negligible. The ground
state fidelity of the Hamiltonian~\eqref{eq:hamil} increases close
to one as $\alpha$ decreases~\cite{sorensen05}.

\begin{figure}
\includegraphics[height=1.16in]{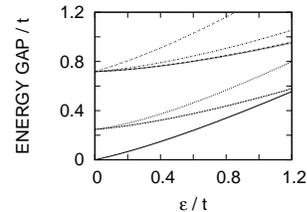}
\caption{Energy gap from the ground state to the nine lowest excited states
for a $4\times4$ lattice in the periodic boundary condition in the
presence of energy shift $-\epsilon\ket{1}\bra{1}$ applied at two
sites chosen evenly. The filling factor is $\nu=1/2$.}
\label{fig:gap}
\end{figure}

In experiments, the ground state could be prepared by the
adiabatic transformation, in a similar manner as in Ref.~\cite{sorensen05}.
To show this, we consider the above $4\times4$ lattice system and
deliberately add an energy shift $-\epsilon[\left(\ket{1}\bra{1}\right)_{0,0}+\left(\ket{1}\bra{1}\right)_{2,2}]$,
which can be done in experiments by applying lasers at those sites
to induce ac Stark shifts. When the energy shift $\epsilon$ is sufficiently
large, the ground state is simply $\ket{1}_{0,0}\ket{1}_{2,2}$ with
all other atoms in state $\ket0$. In FIG.~\eqref{fig:gap}, we plot
the energy gap from the ground state to the nine lowest excited states
with respect to the amount of the energy shift $\epsilon$. The degeneracy
of the ground state in the absence of the energy shift is due to the
ambiguity of the center of mass in the torus geometry, which disappears
in the plane geometry~\cite{haldane85a}. Aside from this degeneracy,
the excited states have finite energy gaps which allow an adiabatic
transformation. From this figure, it is apparent that the Laughlin
ground state can be prepared by the following procedure: (1) Prepare
the atoms in state $\ket{1}$ at sites chosen evenly in agreement
with the filling factor $\nu$, with all other atoms prepared in state
$\ket{0}$. Initially all lasers are turned off; (2) Apply lasers
at the chosen sites to induce an ac Stark shift $-\epsilon\ket{1}\bra{1}$,
with $\epsilon$ chosen moderately, e.g., as the desired value of
$t$. This energy shift does not change the atomic state; (3) Gradually
increase the Rabi frequencies $\Omega^{X}$ and $\Omega^{Y}$ to reach
the desired value of $t$; (4) Gradually decrease the energy shift
$\epsilon$ to zero.

The quasiexcitation of the Laughlin state is
generated when one magnetic flux quantum is adiabatically inserted
through an infinitely thin solenoid piercing the 2D plane~\cite{laughlin83}.
In the present system, we can choose the position of the quasiexcitation
inside a lattice cell. Recalling that the vector potential outside
a solenoid is given by $\vec{A}_{s}=\Phi_{0}/2\pi r\hat{\varphi}$,
where $r$ is the distance from the solenoid and $\hat{\varphi}$
is the azimuthal vector, the effect of the solenoid
can be easily reflected in the phases of Eq.~\eqref{eq:generalphase}. Generation
of the Laughlin state and the existence of the fractionally charged
quasiexcitation (in the present case, fractionally excited bosons)
could be examined by directly measuring the individual atoms: for
example, by measuring the pair correlation functions ~\cite{haldane85c}.
Before the measurement, one may turn off all lasers so as to isolate
the state from further evolution and decoherence.

Although the atomic excitation is highly suppressed, the atomic spontaneous
decay is yet a prominent source of decoherence. If we denote by $\gamma$
the spontaneous decay rate of an atom, the effective decay rate of
the whole system due to the atomic decay is estimated as $N_{b}\gamma(\Omega/\Delta)^{2}$,
where $N_{b}$ denotes the total number of bosons in the system (we
omit superscript $X$ or $Y$ for simplicity). On the other hand,
the energy scale $t$ in the Hamiltonian is given by $J(\Omega/g)^{2}$.
In view of the condition $\Delta\gg g$, the former is still much
smaller than the latter for moderate $N_{b}$ if we assume $\gamma\lesssim J$.
However, since the excitation gap is smaller than $t$, the attainable
system size would be restricted in the experimental realization. Although
the effective decay rate is decreased by increasing $\Delta$, this
in turn requires more stronger atom-cavity coupling rate $g$ so as
to satisfy the condition~\eqref{eq:cond}.

Finally, we stress the point that the asymmetric geometry introduced
in FIG.~\ref{fig:cavity} is not essential. That is, when
the two orthogonal cavity modes have the same resonant frequency,
one can also obtain the Hamiltonian~\eqref{eq:final} in the following
way: we apply lasers with the same frequency, say $\omega_{1}$, in
every second row so that they produce the spin exchange to the $\hat{x}$
direction, while applying lasers with a different frequency $\omega_{2}$ in the other rows, which also produce the spin exchange to the $\hat{x}$
direction. If we choose those frequencies so that $\abs{\omega_{1}-\omega_{2}}\sim\delta^{\mu}$,
they do not produce the spin exchange to the $\hat{y}$ direction.
In the same manner, we apply lasers with frequencies $\omega_{3}$
and $\omega_{4}$ in every second column to produce the spin exchange
to the $\hat{y}$ direction. By choosing those four frequencies to
be sufficiently detuned, we can adjust the associated phases independently
for each pair of the spin exchange.

We thank R. Fazio and B. Hessmo for helpful discussions. This work was supported by the Korea Research Foundation Grant (KRF-2007-357-C00016) funded by the Korean Government (MOEHRD), and by the National Research Foundation \& Ministry of Education, Singapore. SB thanks the Engineering and Physical Sciences Research Council (EPSRC) UK for an Advanced Research Fellowship and for support through the Quantum Information Processing IRC (GR/S82176/01) and the Royal Society and the Wolfson Foundation.

\end{document}